\documentclass[prd,twocolumn,superscriptaddress,showpacs,nofootinbib,preprintnumbers]{revtex4}

\usepackage{amsmath}
\usepackage{amsfonts}
\usepackage{graphicx}
\usepackage{dcolumn}
\usepackage{hyperref}

\newcommand{\be}{\begin{equation}}
\newcommand{\ee}{\end{equation}}
\newcommand{\bea}{\begin{eqnarray}}
\newcommand{\eea}{\end{eqnarray}}
\newcommand{\beaa}{\begin{eqnarray*}}
\newcommand{\eeaa}{\end{eqnarray*}}

\newcommand{\nn}{\nonumber \\}

\newcommand{\e}{{\rm e}}

\begin{document}

\tolerance=5000

\title{Dark Energy: the equation of state description versus scalar-tensor or
 modified gravity}

\author{S. Capozziello}
\thanks{Electronic address: capozziello@na.infn.it}
\affiliation{Dipartimento di Scienze Fisiche, Universit\`{a} di
Napoli ``Federico II'' and INFN, Sez. di Napoli, Compl. Univ.
Monte S. Angelo, Edificio N, Via Cinthia, I-80126, Napoli, Italy}
\author{S. Nojiri}
\thanks{Electronic address: snojiri@yukawa.kyoto-u.ac.jp, \\ nojiri@cc.nda.ac.jp},
\affiliation{Department of Applied Physics, National Defence
Academy, Hashirimizu Yokosuka 239-8686, Japan}
\author{S.D. Odintsov}
\thanks{Electronic address: odintsov@ieec.uab.es also at TSPU, Tomsk,
Russia} \affiliation{
Instituci\`o Catalana de Recerca i Estudis
Avan\c{c}ats (ICREA)  and Institut de Ciencies de l'Espai
(IEEC-CSIC),
Campus UAB, Facultad Ciencies, Torre C5-Par-2a pl,
E-08193 Bellaterra (Barcelona), Spain
}

\begin{abstract}

Dark energy dynamics of the universe can be achieved by equivalent
mathematical descriptions taking into account generalized fluid
equations of state in General Relativity, scalar-tensor theories
or modified $F(R)$ gravity in Einstein or Jordan frames. The
corresponding technique transforming equation of state description
to scalar-tensor or modified gravity is explicitly presented.
 We show
that such equivalent pictures can be discriminated  by matching
solutions with data capable of selecting the true physical frame.

\end{abstract}


\maketitle

\noindent 1. According to recent astrophysical data, our universe
is dominated by a mysterious form of Dark Energy (DE), for a
review and  a list of refs., see \cite{padmanabhan},  which counts
to about 70$\%$ of total mass-energy density. As a result, the
universe expansion is accelerating. In terms of constant equation
of state (EoS) parameterization,  observational data indicate that
constant EoS parameter is roughly equal to $-1$. In other words,
the accelerating universe could be in the cosmological constant
($w=-1$), quintessence ($-1<w<-1/3$) or phantom era ($w<-1$).
Without taking into account the so-called "first and second
coincidence problems" (for a recent discussion, see \cite{cai})
the fundamental problem is to select the (correct) EoS for the
observed universe consistently related to the early epochs.
 Even considering (perfect/imperfect) ideal
fluid  description, there are various possibilities: constant EoS,
time-dependent EoS, complicated (explicit or implicit) EoS
functional dependence of the pressure from  energy density (and
time), inhomogeneous EoS, etc. The situation is even more
complicated since several proposals for DE (from scalars to
string-inspired gravity\cite{sasaki}) exist.

In the present Letter, we develop a technique  by which it is
possible, whatever  the ideal fluid EoS description is, to
transform such a fluid in a scalar-tensor theory taking into
account the "same" FRW scale factor. However, the process can be
reversed. Subsequently, the scalar-tensor theory can be
represented as a modified gravity theory (without scalar field)
with the same scale factor in Jordan or Einstein frames
conformally related. Of course, these three descriptions, leading
to the same FRW dynamics, differ in various respects (for
instance, the Newton law is different, quantum versions of such
theories are not equivalent, nucleosynthesis and LSS can be
achieved in different ways, etc). The proposal is to discriminate
among the three approaches to DE considering  observational data:
in this sense,  the "true" selection of mathematically equivalent
descriptions is operated at the solution level spanning as much as
possible wide ranges of cosmological parameters like the redshift
$z$.

\noindent 2. Let us start from the following action: \be
\label{k1} S=\int d^4 x \sqrt{-g}\left\{\frac{1}{2\kappa^2}R -
\frac{1}{2}\omega(\phi)\partial_\mu \phi
\partial^\mu \phi  - V(\phi)\right\}\ .
\ee Here $\omega(\phi)$ and $V(\phi)$ are functions of the scalar
field $\phi$. The function $\omega(\phi)$ actually may be chosen
to be equal to 1 or -1 as it is shown below. Its possible role is
to realize transitions between deceleration/acceleration phases or
non-phantom and phantom phases \cite{CNO}.
 Let us now assume a spatially-flat FRW metric:
\be \label{k2} ds^2 = - dt^2 + a(t)^2 \sum_{i=1}^3
\left(dx^i\right)^2\ . \ee and that the scalar field $\phi$ only
depends on the time coordinate $t$. Then the FRW equations are
given by \be \label{k3} \frac{3}{\kappa^2}H^2 = \rho\ ,\quad -
\frac{2}{\kappa^2}\dot H= p + \rho\ . \ee Here the energy density
$\rho$ and the pressure $p$ are \be \label{k4} \rho =
\frac{1}{2}\omega(\phi){\dot \phi}^2 + V(\phi)\ ,\quad p =
\frac{1}{2}\omega(\phi){\dot \phi}^2 - V(\phi)\ . \ee Combining
(\ref{k3}) and (\ref{k4}), one obtains \be \label{k5} \omega(\phi)
{\dot \phi}^2 = - \frac{2}{\kappa^2}\dot H\ ,\quad
V(\phi)=\frac{1}{\kappa^2}\left(3H^2 + \dot H\right)\ . \ee The
interesting case is that $\omega(\phi)$ and $V(\phi)$ are defined
in terms of a single function $f(\phi)$ as \be \label{k6}
\omega(\phi)=- \frac{2}{\kappa^2}f'(\phi)\ ,\quad
V(\phi)=\frac{1}{\kappa^2}\left(3f(\phi)^2 + f'(\phi)\right)\ .
\ee Hence, the following solution are obtained \be \label{k7}
\phi=t\ ,\quad H=f(t)\ . \ee One can check that the solution
(\ref{k7}) satisfies the scalar-field equation: \be \label{k8}
0=\omega(\phi)\ddot \phi + \frac{1}{2}\omega'(\phi){\dot\phi}^2 +
3H\omega(\phi)\dot\phi + V'(\phi)\ . \ee Then {\it any} cosmology
defined by $H=f(t)$ in (\ref{k7}) can be realized by (\ref{k6}).

Since we can always redefine the scalar field $\phi$ as $\phi\to
F(\phi)$ by an arbitrary function $F(\phi)$, we can choose the
scalar field to be a time coordinate: $\phi=t$ as in (\ref{k7}).
Using (\ref{k4}), one finds \bea \label{kk1} \rho&=&
\frac{3}{\kappa^2}f(\phi)^2\ ,\nn p&=&
-\frac{3}{\kappa^2}f(\phi)^2 - \frac{2}{\kappa^2}f'(\phi)\ . \eea
Since $\phi=f^{-1}\left(\kappa\sqrt{\rho/3}\right)$, we reobtain
the equation of the state (EoS): \be \label{SN1} p=-\rho -
\frac{2}{\kappa^2}f'\left(f^{-1}\left(\kappa\sqrt{\frac{\rho}{3}}\right)\right)\
, \ee which contains all the cases where the EoS is given by
$p=w(\rho)\rho$ (for a recent study on the acceleration of the
universe in the EoS descrition, see \cite{EoS,sami} and references
therein). Furthermore, since $f^{-1}$ could be always a
single-valued function, Eq.(\ref{SN1}) contains more general EoS
given by \be \label{SN1b} 0=F\left(\rho,p\right)\ . \ee
Conversely, if an EoS is given by (\ref{SN1b}), since $\rho$ and
$p$ are given by (\ref{kk1}), the corresponding $f(\phi)$ can be
obtained by solving the following differential equation: \be
\label{ff1} F\left(\frac{3}{\kappa^2}f(\phi)^2,
-\frac{3}{\kappa^2}f(\phi)^2
 - \frac{2}{\kappa^2}\frac{df(\phi)}{d\phi}\right)=0\ .
\ee If we define a new field $\varphi$ as \be \label{f5}
\varphi=\int d\phi \sqrt{\left|\omega(\phi)\right|}\ , \ee the
action (\ref{k1}) can be rewritten as \be \label{f6} S=\int d^4 x
\sqrt{-g}\left\{\frac{1}{2\kappa^2}R \mp \frac{1}{2}\partial_\mu
\varphi
\partial^\mu \varphi  - \tilde V(\varphi)\right\}\ .
\ee The sign in front of the kinetic term depends on the sign of
$\omega(\phi)$. If the sign of $\omega$ and then the sign of $\dot
H$ is positive (negative), the sign of the kinetic term is $-$
($+$). Therefore, in the phantom phase, the sign is always $+$ and
in the non-phantom phase, always $-$. One assumes $\phi$ can be
solved with respect to $\varphi$: $\phi=\phi(\varphi)$. Then the
potential $\tilde V(\varphi)$ is given by $\tilde V(\varphi)\equiv
V\left(\phi(\varphi)\right)$. Since $\tilde V(\varphi)$ could be
uniquely determined, there is one to one correspondence between
$H$ and $\tilde V(\varphi)$.

By the variation over $\varphi$ in the action (\ref{f6}), in the
FRW metric, the scalar-field equation follows \be \label{ec1}
0=\pm \ddot \varphi + 3H\dot\varphi + {\tilde V}'(\varphi)\ . \ee
Since the energy density and the pressure is now given by \be
\label{ec2} \rho = \pm \frac{1}{2}{\dot \varphi}^2 + \tilde
V(\varphi)\ ,\quad p = \pm\frac{1}{2}{\dot \varphi}^2 - \tilde
V(\varphi)\ , \ee the conservation of the energy can be obtained
by using (\ref{ec1}): \bea \label{ec3} && \dot \rho + 3H
\left(\rho + p\right) \nn &=& \pm \dot\varphi \ddot\varphi +
\tilde V'(\varphi)\dot\varphi + 3H {\dot \varphi}^2  \nn &=&
\dot\varphi \left(\pm \ddot \varphi + 3H\dot\varphi + {\tilde
V}'(\varphi)\right) \nn &=&0\ . \eea Then one can start either
from the EoS ideal fluid description or from the scalar-tensor
theory (\ref{f6}) description: the emerging cosmology is the same.

\noindent 3. Let us consider several examples. As first example,
the simplest case, we take into account a dust model with $p=0$.
Since \be \label{ff1b} -\frac{3}{\kappa^2}f(\phi)^2
 - \frac{2}{\kappa^2}\frac{df(\phi)}{d\phi}=0\ ,
\ee one gets \be \label{ff1c} f(\phi)=\frac{2}{3\phi}\ , \ee which
gives \bea \label{fffo}
\varphi&=&\frac{2}{\kappa\sqrt{3}}\ln\frac{\phi}{\phi_0}\ ,\nn
\tilde V(\varphi)&=&V_0 \e^{-\kappa\varphi\sqrt{3}}\ , \quad V_0
\equiv \frac{2}{3\kappa^2 \phi_0^2}\ . \eea In the case where the
EoS parameter $w=p/\rho$ is constant,  this function is \be
\label{ff1d} f(\phi)=\frac{2}{3(1+w)\phi}\ . \ee Hence, \bea
\label{fff1}
\varphi&=&\frac{2}{\kappa\sqrt{3\left|1+w\right|}}\ln\frac{\phi}{\phi_0}\
,\nn \tilde V(\varphi)&=&V_0
\e^{-\kappa\varphi\sqrt{3\left|1+w\right|}}\ , \nn V_0 &\equiv&
\frac{2(1-w)}{3(1+w)^2\kappa^2 \phi_0^2}\ . \eea Here $\phi_0$ is
an integration constant. Eq.(\ref{ff1d}) gives \be \label{H1}
H=\frac{2}{3(1+w)t}\ , \ee when $w>-1$ and by shifting $t\to t -
t_s$ and considering the case $t<t_s$, we obtain \be \label{H2}
H=-\frac{2}{3(1+w)\left(t_s - t\right)}\ . \ee Eq.(\ref{H2})
expresses the accelerating expansion of the universe. In phantom
case $w<-1$, the sign in front of  the kinetic term in (\ref{f6})
is $+$.

We have to note, however, that the general solution for
(\ref{fff1}) is not restricted to the solution corresponding to a
constant $w$ \cite{russo,alexei}. Let us define new variables $u$,
$v$ and new time coordinate $\tau$ by \bea \label{r1} &&
a=\e^{\frac{v+u}{3}}\ ,\quad \varphi =\frac{2(v-u)}{\sqrt{3}}\
,\nn && d\tau = dt\sqrt{3V_0 \over 8}\e^{-{2(v-u) \over
\varphi_0\sqrt{3\gamma}}}\ ,\nn && \varphi_0\equiv
\frac{2}{\kappa\sqrt{3\left|1+w\right|}}\ . \eea The Hamiltonian
constraint and other equation reduce to \bea \label{r2} &&
\frac{dv}{d\tau}\frac{du}{d\tau} =1\ ,\nn && \frac{d^2 {\cal
U}}{d\tau^2}=\left(1-{\bar \alpha}^2\right){\cal U}\ ,\quad
\frac{d^2 {\cal V}}{d\tau^2}=\left(1-{\bar \alpha}^2\right){\cal
V}\ . \eea Here \be \label{r3} {\cal V}\equiv \e^{(1 - {\bar
\alpha})v}\ ,\quad {\cal U}\equiv \e^{(1+{\bar \alpha})u}\ , \quad
{\bar \alpha}\equiv \frac{2}{\kappa\varphi_0\sqrt{3}} \ee When
$\bar\alpha^2<1$, the general solution of (\ref{r2}) is \bea
\label{r4} {\cal U}&=&u_0\cosh \left(\tau\sqrt{1 -
\bar\alpha^2}\right) + u_1\sinh \left(\tau\sqrt{1 -
\bar\alpha^2}\right)\ ,\nn {\cal V} &=& v_0\left(u_1\cosh
\left(\tau\sqrt{1 - \bar\alpha^2}\right) \right. \nn && \left. +
u_0\sinh \left(\tau\sqrt{1 - \bar\alpha^2}\right)\right)\ , \eea
and when $\bar\alpha^2<1$, \bea \label{r5} {\cal U}&=&\hat u_0\cos
\left(\tau\sqrt{1 - \bar\alpha^2}\right) + \hat u_1\sin
\left(\tau\sqrt{1 - \bar\alpha^2}\right)\ ,\nn {\cal V} &=&\hat
v_0\left(\hat u_1\cosh \left(\tau\sqrt{1 - \bar\alpha^2}\right)
\right. \nn && \left. - \hat u_0\sinh \left(\tau\sqrt{1 -
\bar\alpha^2}\right)\right)\ , \eea Here $u_0$, $u_1$, $v_0$,
$\hat u_0$, $\hat u_1$, $\hat v_0$ are constants determined by a
proper initial conditions.

In the case of a Chaplygin gas, whose EoS is given by \be
\label{ff2} p\rho=-A\ , \ee with positive constant $A$, $f(\phi)$
can be obtained from the following algebraic equation \be
\label{ff3} \frac{i}{6\zeta}\ln \left(\frac{f(\phi) - i\zeta
a}{f(\phi) + i\zeta a}\right) - \frac{1}{6\zeta}\ln
\left(\frac{f(\phi) - \zeta a}{f(\phi) + \zeta a}\right) =\phi\ .
\ee Here \be \label{ff4} \zeta\equiv \e^{i\pi/4}\ ,\quad a^4\equiv
\frac{\kappa^4 A}{9}\ . \ee Let the solution of (\ref{ff3}) be
$f(\phi)=f_C(\phi)$. Then by using (\ref{k6}) and (\ref{f5}), we
may define a new scalar field as \be \label{C1}
\varphi_C=\frac{2}{\kappa}\int d\phi
\sqrt{\left|f_C(\phi)\right|}\ , \ee which may be solved as
$\phi=\phi\left(\varphi_C\right)$. The
 corresponding potential
$\tilde V_C(\varphi_C)$ could be given as \be \label{C2} \tilde
V_C(\varphi_C)=\frac{1}{\kappa^2}\left(3f_C\left(\phi\left(\varphi_C\right)\right)^2
+ f'\left(\phi\left(\varphi_C\right)\right)\right)\ . \ee

As next example we consider the following $f(\phi)$: \be
\label{f0} f(\phi)=h_0\left(\frac{1}{\phi} + \frac{1}{t_s
-\phi}\right)\ , \ee with constant $h_0$, which gives \be
\label{f1} H=h_0\left(\frac{1}{t} + \frac{1}{t_s -t}\right)\ . \ee
This corresponds to the following EoS: \be \label{f2} 0=4\kappa^2
t_s^2 \rho^2 - 9\kappa^2 h_0^2 \left(p + \rho\right)^2 - 48 h_0
t_s \rho\ . \ee In this case, one finds \be \label{t1}
\varphi=\frac{\sqrt{2h_0}}{\kappa}\int d\phi
\sqrt{\left|\frac{1}{\phi^2} - \frac{1}{\left(t_s - \phi
\right)^2}\right|}\ . \ee By solving (\ref{t1}) with respect to
$\varphi$ as $\phi=\phi(\varphi)$, we obtain the corresponding
potential: \bea \label{t2} \tilde
V(\varphi)&=&\frac{1}{\kappa^2}\left\{3h_0^2\left(\frac{1}{\phi(\varphi)}
+ \frac{1}{t_s - \phi(\varphi)} \right)^2 \right. \nn && \left. +
h_0\left(\frac{1}{\phi\left(\varphi\right)^2} - \frac{1}{\left(t_s
- \phi\left(\varphi\right)\right)^2} \right)\right\}\ . \eea Since
\be \label{p1} \dot H= \frac{h_0\left( 2t - t_s\right)t_s}{t^2
\left(t_s -t\right)}\ , \ee when $t<t_s/2$, the universe is in
non-phantom phase ($\dot H<0$) but when $t>t_s/2$, in phantom
phase ($\dot H>0$). While using $\phi$, the transition is smooth
although $\omega(\phi)=0$ at the transition point but the
definition of $\varphi_0$ in (\ref{t1}) has a discontinuity there.
Then the EoS (\ref{f2}) admits the phantom phase but, in such a
phase, the sign of  $\varphi$-kinetic term   in (\ref{f6}) becomes
$+$.

Now one may consider \be \label{f3}
f(\phi)=\left\{\frac{3}{2}\left(1-2\alpha\right)\left(\frac{3}{\kappa^2}\right)^{\alpha
- 1}f_0 \right\}^{\frac{1}{1 - 2\alpha}}\phi^{\frac{1}{1 -
2\alpha}}\ . \ee with constants $f_0$ and $\alpha$. The
corresponding EoS is \be \label{f4} p=-\rho + f_0 \rho^\alpha\ .
\ee By using (\ref{f5}), we find \bea \label{AA1}
\varphi&=&\varphi_0 \phi^{\frac{1-\alpha}{1-2\alpha}}\ ,\nn
\varphi_0&\equiv&\frac{\sqrt{2\left|(1-2\alpha)f_1\right|}}{\left|\alpha\right|\kappa}\
,\nn f_1&\equiv&
\left\{\frac{3}{2}\left(1-2\alpha\right)\left(\frac{3}{\kappa^2}\right)^{\alpha
- 1}f_0 \right\}^{\frac{1}{1 - 2\alpha}}\ . \eea Then the
corresponding potential is given by \be \label{AA2} \tilde
V(\varphi)=\frac{1}{\kappa^2}\left\{ f_1^2
\left(\frac{\varphi}{\varphi_0}\right)^{\frac{2}{1-\alpha}} +
\frac{f_1}{1-2\alpha}
\left(\frac{\varphi}{\varphi_0}\right)^{\frac{2\alpha}{1-\alpha}}
\right\}\ . \ee Eq.(\ref{f3}) gives \be \label{HH1} H=f_1
t^{\frac{1}{1-2\alpha}}\ ,\quad \dot H=\frac{f_1
t^{\frac{2\alpha}{1 - 2\alpha}}}{1-2\alpha}\ . \ee Since $f_1$ in
(\ref{AA2}) is positive, if $\alpha<1/2$, the universe is in the
phantom phase. In order that $f_1$ is real, we also require
$f_0>0$. In the phantom (non-phantom) phase, the sign of the
kinetic term of $\varphi$ in (\ref{f6}) becomes $+$ ($-$).

As further example, we consider the following EoS: \be \label{ff5}
\rho = -p + \frac{AB\rho^{\alpha+\beta}}{A\rho^\alpha +
B\rho^\beta}\ , \ee with constants, $A$, $B$, $\alpha$, $\beta$.
This EoS has been proposed in \cite{NOT}. By solving (\ref{ff1}),
we find that $f(\phi)$ is the solution of the following algebraic
equation: \bea \label{ff6}
&&-\frac{2}{\kappa^2}\left\{\frac{1}{\left(1 -
\frac{\beta}{2}\right)B}
\left(\frac{3}{\kappa^2}\right)^{-\frac{\beta}{2}} f(\phi)^{1 -
\frac{\beta}{2}} \right. \nn && \left. + \frac{1}{\left(1 -
\frac{\alpha}{2}\right)A}
\left(\frac{3}{\kappa^2}\right)^{-\frac{\alpha}{2}} f(\phi)^{1 -
\frac{\alpha}{2}} \right\}=\phi \ . \eea Let the solution of
(\ref{ff6}) be $f(\phi)=f_S(\phi)$. Then one may define a new
scalar field as \be \label{S1} \varphi_S=\frac{2}{\kappa}\int
d\phi \sqrt{\left|f_S(\phi)\right|}\ , \ee which can be solved as
$\phi=\phi\left(\varphi_S\right)$. The
 corresponding potential
$\tilde V_S(\varphi_C)$ is given as \be \label{S2} \tilde
V_S(\varphi_S)=\frac{1}{\kappa^2}\left(3f_S\left(\phi\left(\varphi_S\right)\right)^2
+ f'\left(\phi\left(\varphi_S\right)\right)\right)\ . \ee

As final example, we consider a fluid with a Van der Waals EoS
type \cite{CTTC} \be \label{ff7} p=\frac{\gamma\rho}{1 -
\beta\rho} - \alpha \rho^2\ , \ee where $\alpha$, $\beta$, and
$\gamma$ are constants. The corresponding $f(\phi)$ is the
solution of the following differential equation: \be \label{ff8}
f'(\phi)=-\frac{\frac{3}{2}\gamma f(\phi)^2}{1 -
\frac{3\beta}{\kappa^2}f(\phi)^2} +
\frac{9\alpha}{\kappa^4}f(\phi)^4 - \frac{3}{2} f(\phi)^2\ , \ee
which is not so easy to solve explicitly. For some solution of
(\ref{ff8}) as $f(\phi)=f_{VdW}(\phi)$,  a new scalar field can be
defined  as \be \label{VdW1} \varphi_{VdW}=\frac{2}{\kappa}\int
d\phi \sqrt{\left|f_{VdW}(\phi)\right|}\ , \ee which can be solved
as $\phi=\phi\left(\varphi_{VdW}\right)$. The corresponding
potential $\tilde V_{VdW}(\varphi_{VdW})$ is found to be \bea
\label{VdW2} \tilde
V_{VdW}(\varphi_{VdW})&=&\frac{1}{\kappa^2}\left(3f_{VdW}
\left(\phi\left(\varphi_{VdW}\right)\right)^2 \right. \nn &&
\left. + f'\left(\phi\left(\varphi_{VdW}\right)\right)\right)\ .
\eea Similarly, one can construct, in principle, the
correspondence between any dark energy EoS  and a given
scalar-tensor theory. In other words, any explicit (implicit)
ideal-fluid EoS of the universe, governed by General Relativity,
could be represented as some scalar-tensor theory with specific
potential, and vice-versa.

\noindent 4. Let us now investigate the relations between the
scalar-tensor theory (\ref{k1}) and modified $F(R)$-gravity, whose
Lagrangian density is given by a proper function $F(R)$ of the
scalar curvature $R$. In such a case, the sign in front of the
kinetic term of $\varphi$ in (\ref{f6}) is $-$. We can use the
conformal transformation \be \label{f7} g_{\mu\nu}\to
\e^{\pm\kappa \varphi\sqrt{\frac{2}{3}}}g_{\mu\nu}\ , \ee and make
the kinetic term of $\varphi$ vanish. Hence, one obtains \be
\label{f8} S=\int d^4 x \sqrt{-g}\left\{\frac{\e^{\pm\kappa
\varphi\sqrt{\frac{2}{3}}}}{2\kappa^2}R
 - {\e^{\pm 2\kappa \varphi\sqrt{\frac{2}{3}}}}\tilde V(\varphi)\right\}\ .
\ee The action  (\ref{f8}) is the so-called  "Jordan frame action"
while the action (\ref{f6}) is the "Einstein frame action" due to
either the non-minimal coupling or the standard coupling in front
of the Ricci scalar. Since $\varphi$ is the auxiliary field, one
may cancel out $\varphi$ by using the
 equation of motion:
\be \label{f9} R=\e^{\pm\kappa \varphi\sqrt{\frac{2}{3}}}
\left(4\kappa^2 \tilde V(\varphi) \pm 2\kappa \sqrt{\frac{3}{2}}
\tilde V'(\varphi)\right)\ , \ee which can be solved with respect
to $R$ as $\varphi=\varphi(R)$.
 We can rewrite
the action (\ref{f8}) in the form of $F(R)$-gravity: \bea
\label{f10} S&=&\int d^4 x \sqrt{-g}F(R)\ , \nn F(R) &\equiv&
\frac{\e^{\pm \kappa \varphi(R)\sqrt{\frac{2}{3}}}}{2\kappa^2}R
  - {\e^{\pm 2\kappa \varphi(R)\sqrt{\frac{2}{3}}}}\tilde V\left(\varphi(R)\right)\ .
\eea Note that one can rewrite the scalar-tensor theory (\ref{k1})
or equivalently (\ref{f6}), only when the sign in front of the
kinetic term is $-$ in (\ref{f6}), that is, $\omega(\phi)$ is
positive. In the phantom phase, $\omega(\phi)$ is negative. In
this case, the above procedure to rewrite the phantom
scalar-tensor theory as  $F(R)$-gravity does not work. However, we
should note that, for the metric transformed as in (\ref{f7}),
 even if the universe
in the original metric (corresponding to the Einstein frame) is in
a phantom phase, the universe in the re-scaled metric
(corresponding to the Jordan frame) can be, in general, in a
non-phantom phase.

For example, in the case of (\ref{ff1d}) or (\ref{fff1}), after
the scale transformation (\ref{f7}) and cancelling out $\varphi$,
one obtains the action corresponding to (\ref{fff1}): \be
\label{fff2} S=A\int d^4x \sqrt{-g} R^\alpha\ . \ee Here \bea
\label{fff3} A&\equiv& \left(1 \mp 3\sqrt
\frac{1+w}{2}\right)V_0^{-\frac{1}{1 \mp 3\sqrt{\frac{1+w}{2}}}}
\nn && \times \left\{2\kappa^2\left(2 \mp 3
\sqrt{\frac{1+w}{2}}\right) \right\}^{-1-\frac{1}{1 \mp
3\sqrt{\frac{1+w}{2}}}} \ ,\nn \alpha&\equiv& 1+\frac{1}{1 \mp
3\sqrt{\frac{1+w}{2}}} \ . \eea We have to note that, if we start
from the action (\ref{fff2}), due to the conformal transformation
(\ref{f7}), the behavior of the universe is different from that
given by (\ref{ff1d}), which is $a=a_0 t^{\frac{2}{3(1+w)}}$ with
a constant $a_0$ (of course, also the Newton law, in the weak
energy limit, is modified and interesting results come out as the
fact that flat rotation curves of galaxies could be explained {\it
without} the need of dark matter \cite{PLA}.) By the conformal
transformation (\ref{f7}), the FRW metric (\ref{k2}),
corresponding to the Einstein frame, is transformed as \bea
\label{fff4} && ds^2 \to \nn && d{\tilde s}^2\equiv \e^{\mp \kappa
\varphi\sqrt{\frac{2}{3}}}ds^2 =
\left(\frac{t}{\phi_0}\right)^{\mp\frac{2}{3}\sqrt{\frac{2}{1+w}}}
\nn && \times \left\{- dt^2 + a(t)^2 \sum_{i=1}^3
\left(dx^i\right)^2\right\}\ . \eea The transformed metric can be
regarded as that in the Jordan frame. With a new cosmic time
$\tilde t$ \be \label{fff5} d\tilde t =
\left(\frac{t}{\phi_0}\right)^{\mp
\frac{1}{3}\sqrt{\frac{2}{1+w}}}dt\ , \ee
 $d{\tilde s}^2$  can be written in the FRW form as:
\be \label{fff6} d{\tilde s}^2\ = - d{\tilde t}^2 + \tilde
a(\tilde t)^2 \sum_{i=1}^3 \left(dx^i\right)^2\ . \ee Hence,
$\tilde a(\tilde t)$ behaves as \be \label{fff7} \tilde a(\tilde
t)\sim {\tilde t}^{\frac{\frac{2}{3(1+w)} \mp
\frac{1}{3}\sqrt{\frac{2}{1+w}}} {1 \mp
\frac{1}{3}\sqrt{\frac{2}{1+w}}}}\ . \ee Therefore if we start
from the action (\ref{fff2}), the behavior of the universe is
different from that given by (\ref{ff1d}). The exponent in
(\ref{fff7}) can be defined as $\tilde h_0$: \be \label{hhh1}
\tilde h_0 \equiv \frac{\frac{2}{3(1+w)} \mp
\frac{1}{3}\sqrt{\frac{2}{1+w}}} {1 \mp
\frac{1}{3}\sqrt{\frac{2}{1+w}}}\ . \ee In case of $-$ sign in
(\ref{hhh1}), we find $h_0<0$ when \be \label{hhh2}
-\frac{7}{9}<w<1\ . \ee Then if we shift $t$ as $t\to t - t_s$,
when $t<t_s$, the universe is in a  phantom phase (super
accelerating). In terms of $\alpha$ in the action (\ref{fff2}),
which is defined in (\ref{fff3}), the region in (\ref{hhh2})
corresponds to \be \label{hhh3} \alpha>0\ , \ee which is
consistent with the result in \cite{ANO}\footnote{ In \cite{ANO},
in order that the $R^\alpha$-term dominates compared with the
Einstein-Hilbert term, $n=-\alpha$ is restricted to be $n>-1$.
Then in \cite{ANO}, the condition for the super-accelerating
expansion is $-1/2>n>-1$ or $1/2<\alpha<1$. }. Hence, even if the
universe is not super-accelerating in the original Einstein frame,
it is super-accelerating in the shifted Jordan frame in
(\ref{f7}). This demonstrates that being mathematically
equivalent, the physics in two such frames may be different.

For the model (\ref{f4}), the equation corresponding to (\ref{f9})
has the following form: \bea \label{R1}
R&=&\e^{\kappa\varphi\sqrt{\frac{3}{2}}}\left[ 4\left\{ f_1^2
\left(\frac{\varphi}{\varphi_0}\right)^{\frac{2}{1-\alpha}}
\right.\right. \nn && \left. + \frac{f_1}{1-2\alpha}
\left(\frac{\varphi}{\varphi_0}\right)^{\frac{2\alpha}{1-\alpha}}
\right\} \nn && + \frac{2}{\kappa \varphi_0}\sqrt{\frac{3}{2}}
\left\{ \frac{2f_1^2}{1 -\alpha}
\left(\frac{\varphi}{\varphi_0}\right)^{\frac{1+\alpha}{1-\alpha}}
\right. \nn && \left.\left. + \frac{2\alpha
f_1}{(1-2\alpha)(1-\alpha)}
\left(\frac{\varphi}{\varphi_0}\right)^{-\frac{1-3\alpha}{1-\alpha}}
\right\}\right] \ , \eea which can, in principle, be solved with
respect to $\varphi$ as $\varphi=\varphi(R)$. Then the
$F(R)$-gravity action corresponding to (\ref{f10}) has the
following form: \bea \label{R2} F(R) &=& \frac{\e^{\kappa
\varphi(R)\sqrt{\frac{2}{3}}}}{2\kappa^2}R \nn &&  -
\frac{\e^{2\kappa \varphi(R)\sqrt{\frac{2}{3}}}}{\kappa^2} \left\{
f_1^2
\left(\frac{\varphi(R)}{\varphi_0}\right)^{\frac{2}{1-\alpha}}
\right. \nn && \left. + \frac{f_1}{1-2\alpha}
\left(\frac{\varphi(R)}{\varphi_0}\right)^{\frac{2\alpha}{1-\alpha}}
\right\}\ . \eea

Thus, the explicit examples presented above show that
(canonical/quintessence) scalar-tensor theory may be always mapped
to modified gravity theory with the same FRW dynamics in one of
the frames but the corresponding Newton law is different.

As a generalization of (\ref{k1}), we may consider the following
action: \bea \label{kkk1} S&=&\int d^4 x
\sqrt{-g}\Bigl\{\frac{1}{2\kappa^2}R -
\frac{1}{2}\omega(\phi)\partial_\mu \phi
\partial^\mu \phi \nn
&& + \eta(\phi) R - V(\phi)\Bigr\}\ . \eea Here $\eta(\phi)$ is a
function of $\phi$. The late-time, accelerating cosmology for the
above theory has been discussed in detail in Ref.\cite{faraoni}
and refs. therein. It is worth noting that the stability
conditions for the above theory, found from perturbation analysis
(see last ref. in \cite{faraoni}) confirms such stability
conditions, earlier defined  in ref.\cite{guido} for equivalent
modified gravity where quantum considerations have been used.

 As before, FRW metric
(\ref{k2}) and $\phi$ only depends on $t$. The explicit
calculation gives \bea \label{kkk2} \rho &=&
\frac{1}{2}\omega(\phi){\dot \phi}^2 + V(\phi) - 6H^2 \eta(\phi) -
6H\eta'(\phi)\dot\phi\ ,\nn p &=& \frac{1}{2}\omega(\phi){\dot
\phi}^2 - V(\phi) + 2\left(2\dot H + 3H^2\right)\eta(\phi) \nn &&
+ 2 \eta''(\phi){\dot\phi}^2 + 2\eta'(\phi)\ddot\phi + 6H
\eta'(\phi)\dot\phi\ . \eea When $\phi=t$, Eqs.(\ref{kkk2}) reduce
to \bea \label{kkk3} \rho &=& \frac{1}{2}\omega(\phi) + V(\phi) -
6H^2 \eta(\phi) - 6H\eta'(\phi)\ ,\nn p &=&
\frac{1}{2}\omega(\phi) - V(\phi) + 2\left(2\dot H +
3H^2\right)\eta(\phi) \nn && + 2 \eta''(\phi) + 6H \eta'(\phi)\ .
\eea Deleting $\phi$ from  Eqs.(\ref{kkk3}), we obtain the
inhomogeneous EoS description recently introduced  in \cite{inh}:
\be \label{KKK4} F(\rho,p,H,\dot H)=0\ . \ee For specific cases,
such inhomogeneous EoS description is equivalent to EoS
description with time-dependent bulk viscosity\cite{brevik}.

\begin{table}
\begin{center}
\begin{tabular}{|ccccc|} \hline
  $EoS$ & $\longleftrightarrow$ & ${\cal L}_{ST}$ & $\longleftrightarrow$ &
     ${\cal L}_{F(R)}$ \\
  $\updownarrow$ &  & $\updownarrow$ &  & $\updownarrow$ \\
  Einstein Eqs. & $\longleftrightarrow$ & ST Eqs. & $\longleftrightarrow$ & $F(R)$ Eqs. \\
  $\updownarrow$ &  & $\updownarrow$ &  & $\updownarrow$ \\
  E-frame Sol. & $\longleftrightarrow$ & E-frame Sol.+$\phi$ & $\longleftrightarrow$ &
  J-frame Sol. \\ \hline
\end{tabular}
\end{center}
\caption{Summary of the three approaches (Eos, Scalar-Tensor and
$F(R)$) equivalent at level of Lagrangians, field equations and
solutions. The solutions are in the Einstein frame for EoS and
ST-gravity while they are found in the Jordan frame for
$F(R)$-gravity. }
\end{table}

\noindent 5. The previous results can be summarized in the Table
I, where the mathematical equivalence of the three approaches is
shown. In principle, the  physical frame (Einstein or Jordan) and
the "true" physical approach (EoS, ST or $F(R)$) is "selected" by
experimental data and observations which should show if the
effective EoS is not simply that of a perfect fluid, if a
quintessence scalar field really exists or if a $F(R)$-theory of
gravity has to be invoked to solve the shortcomings of General
Relativity (in this sense, also Dark Matter and Dark Energy could
be considered as "shortcomings" of General Relativity since no
definitive proof of their existence has been given, up to now, at
some fundamental level). Going to  cosmology, solutions have to be
matched with observations by using the redshift $z$ as the natural
time variable for the Hubble parameter, i.e.

\be H(z)=f(z,\dot{z})=-\frac{\dot{z}}{z+1}\,. \ee  Interesting
ranges for $z$  are: $100< z < 1000$ for early universe (CMBR
data), $10 < z < 100$ (LSS), $0 < z < 10$ (SNeIa, radio-galaxies,
etc.). The method consists in building up a reasonable patchwork
of data coming from different epochs and then
 matching them with the "same" cosmological solution ranging, in principle, from
inflation to present accelerated era. Considering the  example in
previous section, the "mathematical equivalence"  between Jordan
and Einstein frame of  cosmological solutions in scalar-tensor or
in $F(R)$-gravity can be removed if, in a certain redshift range,
data confirm or not super-acceleration: the physical frame (and
then the physical theory) is the one in agreement with data.
Specifically, the method can be outlined as follows. In order to
constrain the parameters characterizing the cosmological solution,
we have to maximize the following likelihood function\,:

\begin{equation}
{\cal{L}} \propto \exp{\left [ - \frac{\chi^2({\bf p})}{2} \right
]} \label{eq: deflike}
\end{equation}
where {\bf p} are the parameters of the cosmological solution and
the $\chi^2$ merit function is defined as\,:

\begin{eqnarray}
\chi^2({\bf p}) & = & \sum_{i = 1}^{N}{\left [ \frac{y^{th}(z_i,
{\bf p}) - y_i^{obs}}{\sigma_i} \right ]^2}
\nonumber \\
~ & + & \displaystyle{\left [ \frac{{\cal{R}}({\bf p}) -
1.716}{0.062} \right ]^2} + \displaystyle{\left [
\frac{{\cal{A}}({\bf p}) - 0.469}{0.017} \right ]^2}  \ .
\label{eq: defchi}\
\end{eqnarray}
Terms entering Eq.(\ref{eq: defchi}) can be characterized as
follows. The dimensionless coordinate distances $y$ to objects at
redshifts $z$ are considered in the first term. They are defined
as\,:

\begin{equation}
y(z) = \int_{0}^{z}{\frac{dz'}{E(z')}} \label{eq: defy}
\end{equation}
where $E(z)=H(z)/H_0$ is a normalized Hubble parameter. This is
the quantity which allows to compare the theoretical results with
data. The function $y$ is related to the luminosity distance $D_L
= (1 + z) r(z)$ or, equivalently, to the distance modulus $\mu$.
A sample of data on $y(z)$ for the 157 SNeIa in the Riess et al.
\cite{Riess04} Gold dataset and 20 radio-galaxies from
\cite{RGdata} is in \cite{DD04}. These authors fit with good
accuracy the linear Hubble law at low redshift ($z < 0.1$)
obtaining the Hubble dimensionless parameter $h = 0.664 {\pm}
0.008\,.$  Such a number can be consistently taken into account at
low redshift.  This value  is in agreement with $H_0 = 72 {\pm} 8
\ {\rm km \ s^{-1} \ Mpc^{-1}}$ given by the HST Key project
\cite{Freedman} based on the local distance ladder and estimates
coming from  time delays in multiply imaged quasars \cite{H0lens}
and Sunyaev\,-\,Zel'dovich effect in X\,-\,ray emitting clusters
\cite{H0SZ}. The second term in Eq.(\ref{eq: defchi}) allows to
extend the $z$-range to probe $y(z)$ up to the last scattering
surface $(z\geq 1000)$.  The {\it shift parameter}
\cite{WM04,WT04} $ {\cal R} \equiv \sqrt{\Omega_M} y(z_{ls}) $ can
be determined from the CMBR anisotropy spectrum, where $z_{ls}$ is
the redshift of the last scattering surface which can be
approximated as  $ z_{ls} = 1048 \left ( 1 + 0.00124
\omega_b^{-0.738} \right ) \left ( 1 + g_1 \omega_M^{g_2} \right )
$ with $\omega_i = \Omega_i h^2$ (with $i = b, M$ for baryons and
total matter respectively) and $(g_1, g_2)$ given in \cite{HS96}.
The parameter $\omega_b$ is constrained by the baryogenesis
calculations contrasted to the observed abundances of primordial
elements. Using this method, the value $ \omega_b = 0.0214 {\pm}
0.0020  $  is found \cite{Kirk}. In any case, it is worth noting
that the exact value of $z_{ls}$ has a negligible impact on the
results and setting $z_{ls} = 1100$ does not change constraints
and priors on the other  parameters of the given model. The third
term in the function $\chi^2$ takes into account  the {\it
acoustic peak} of the large scale correlation function at $100 \
h^{-1} \ {\rm Mpc}$ separation, detected by
 using  46748 luminous red galaxies (LRG)
selected from the SDSS Main Sample \cite{Eis05,SDSSMain}. The
quantity

\begin{equation}
{\cal{A}} = \frac{\sqrt{\Omega_M}}{z_{LRG}} \left [
\frac{z_{LRG}}{E(z_{LRG})} y^2(z_{LRG}) \right ]^{1/3} \label{eq:
defapar}
\end{equation}
is related to the position of acoustic peak where $z_{LRG} = 0.35$
is the effective redshift of the above sample. The parameter
${\cal{A}}$ depends  on the dimensionless coordinate distance (and
thus on the integrated expansion rate),  on $\Omega_M$ and $E(z)$.
This dependence removes some of the degeneracies intrinsic in
distance fitting methods. Due to this reason, it is particularly
interesting to include ${\cal{A}}$ as a further constraint on the
model parameters using its measured value  $ {\cal{A}} = 0.469
{\pm} 0.017  $ reported in \cite{Eis05}. Note that, although
similar to the usual $\chi^2$ introduced in statistics, the
reduced $\chi^2$ (i.e., the ratio between the $\chi^2$ and the
number of degrees of freedom) is not forced to be 1 for the best
fit model because of the presence of the priors on ${\cal{R}}$ and
${\cal{A}}$ and since the uncertainties $\sigma_i$ are not
Gaussian distributed, but take care of both statistical errors and
systematic uncertainties. With the definition (\ref{eq: deflike})
of the likelihood function, the best fit model parameters are
those that maximize ${\cal{L}}({\bf p})$.

\noindent 6. Using the method sketched above, the models studied
here can be constrained and selected by  observations. From an
observational point of view, inhomogeneous EoS with further terms
 added to $p=-\rho$ are preferable for the following reasons. Several
  evidences indicates that $\Lambda$CDM
$(p=-\rho)$ is the cosmological scenario able to realistically
describe  the today observed universe. Any evolutionary model
passing from deceleration (dark matter dominance) to acceleration
(dark energy dominance) should consistently reproduce, based on
the  today available observations, such a scenario.  Adding terms
in Hubble parameter and its derivative to the $\Lambda$ EoS
allows, in any case, a comparison with standard matter parameters
($w_M$ and $\Omega_M$) which are directly observable by
astrophysical standard methods. Then the number of arbitrary
choices (for example,  fixing priors) is not so large. On the
other hand, implicit EoS, of the general form in Eq.(\ref{SN1b}),
needs several arbitrary choices which could result completely
inconsistent to further and more refined observations. For
example,  phantom-like regimes, which could result consistent with
observations at large $z$ (far distances and early universe),
could be improperly discarded imposing arbitrary constraints on
$q_0$ at present epoch. Due to these reasons, from an
observational point of view, it is preferable to study models
which imply corrections to the $\Lambda$ EoS rather than giving
EoS in implicit form  as shown, for example, in
Refs.\cite{CTTC,inh}. However, from a theoretical viewpoint, this
is not a  definitive enough argument in favor of EoS picture.

\begin{figure}
\centering \resizebox{8 cm}{!}{\includegraphics{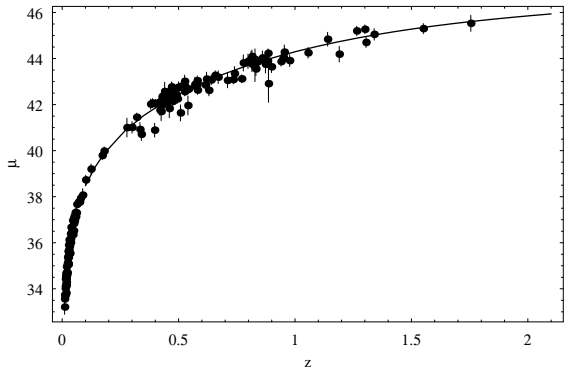}}
\caption{The distance moduli  $\mu$ for the $\Lambda$CDM and the
exponential potential models, obtained from the best fit
parameters of Table II, are compared with SNeIa data in
\cite{Riess04}. The two curves practically  coincide for $z\leq
2$.} \label{fig1}
\end{figure}

In order to give a significant  example, let us compare, from an
observational standpoint,  the $\Lambda$CDM model with the
exponential potential model of scalar-tensor theory. With respect
to the arguments presented in this paper, these are paradigmatic
examples since they can be always conformally transformed from
Jordan to Einstein frame and vice-versa,  corresponding
$F(R)$-models can be  recovered in any case, and, finally,
observational parameters are always referred to the $\Lambda$CDM
model. The constant $w$-case is a particular solution of
Eq.(\ref{fff1}) as discussed above. As said, we can constrain the
cosmological parameters considering the likelihood function
(\ref{eq: deflike}) and the method sketched above. Using the
values of cosmological parameters derived from most popular
datasets \cite{WMAP,CBI}, the two models seem to coincide (Table
II) and are practically the same at low redshifts (see
Fig.(\ref{fig1})). In order to improve the result, we have taken
into account also the reionization redshift $z_{re}$ and the
spectral index of scalar fluctuations $n_s$.

\begin{table}
\begin{center}
\begin{tabular}{|c|c|c|c|c|c|c|}
\hline & \multicolumn{3}{|c|}{$\Lambda$CDM $-\log {\cal L
}=765.3$} & \multicolumn{3}{|c|}{Exp$\varphi$ $-\log {\cal L}=767.3$} \\
\hline
par. & best fit & lower & upper & best fit & lower & upper \\
\hline $\Omega _{b}h^{2}$ & $0.0226$ & $0.0206$ & $0.0256$ &
$0.023$ & $0.0213$ & $ 0.0266$ \\ \hline $\Omega _{DM}h^{2}$ &
$0.120$ & $0.103$ & $0.139$ & $0.110$ & $0.094$ & $ 0.134$ \\
\hline $n_{s}$ & $0.960$ & $0.914$ & $1.05$ & $0.948$ & $0.905$ &
$1.04$ \\ \hline $\Omega
_{M}$ & $0.298$ & $0.222$ & $0.379$ & $0.298$ & $0.232$ & $0.383$ \\
\hline $z_{re}$ & $12.1$ & $2.57$ & $24.0$ & $12.6$ & $2.50$ &
$23.6$ \\ \hline $h$ & $0.692$ & $0.643$ & $0.770$ & $0.669$ &
$0.628$ & $0.729$ \\ \hline
\end{tabular}
\end{center}
\caption{Best fit parameters comparing $\Lambda$CDM and
exponential potential models using the  WMAP \cite{WMAP} and CBI
\cite{CBI} data. ${\cal L}$ is the likelihood function defined in
the text. The lower and upper limits of the parameter values are
the extremal points of the $6$-dimensional confidence region. At a
first look, the two models seem compatible.}
\end{table}

\begin{figure}
\centering \resizebox{8 cm}{!}{\includegraphics{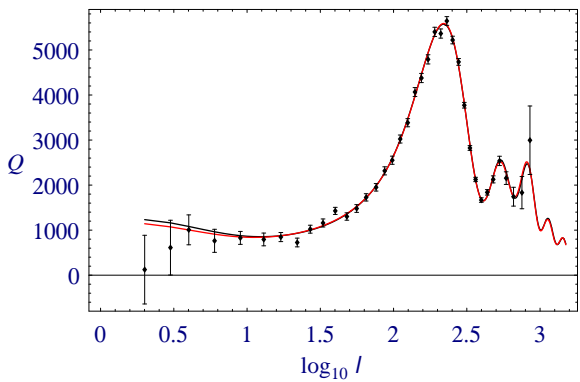}}
\caption{The CMBR angular power spectrum $Q \equiv
l(l+1)C_l/2\protect\pi$ for the two models, obtained with CAMB
codes \cite{CAMB} from the best fit parameters of Table II. The
two curves do not  coincide  for small $l$'s, where the
exponential potential gives higher values. We have to note that we
are exploring a different range of $z$ with respect to that in
Fig.1.}\label{fig2}
\end{figure}

The situation slightly changes if CMBR angular power spectrum is
used. In this case, the set of data is more complete and refined
with respect to the SNeIa ones. From Fig.(\ref{fig2}), it is clear
that models differ for small $l$.

\begin{figure}
\centering \resizebox{8 cm}{!}{\includegraphics{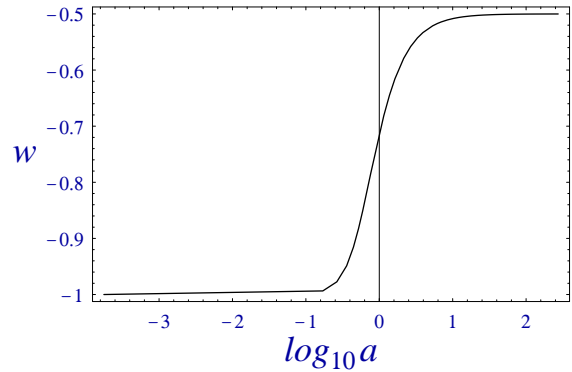}}
\caption{Plot of the scalar-field equation of state versus
${\log}_{10}a$ with the best fit value of $\Omega_M = 0.298$. The
vertical bar marks the today value of scale factor ${\log}_{10}
a_0$. Only with this choice of variables, there is evidence of a
transition from $w \approx -1$ in the past to $w\approx-0.5$ in
the future. This means that other constant values of $w$ can be
generically recovered from scalar-tensor theory, also for
exponential potentials, so that $\Lambda$CDM does not coincides
with exponential models for any $w$ and any $z$.}\label{fig3}
\end{figure}

The differences between the two models are put in evidence if one
chooses suitable variables by which representing the $w$ evolution
in asymptotic regimes, as shown in Fig.(\ref{fig3}). This fact
agrees with previous considerations by which inhomogeneous EoS
could realistically represent the today observed cosmological
scenario allowing also to achieve the early and the late evolution
of the universe.

In conclusion, larger and more detailed samples of data  than
those today available are needed to fit solutions in wide ranges
of $z$.  Only in this situation, the true physical frame could be
univocally selected.

\end{document}